\documentclass[10pt, conference, compsocconf]{IEEEtran}
\IEEEoverridecommandlockouts
\usepackage{cite}
\usepackage{amsmath,amssymb,amsfonts}
\usepackage{algorithmic}
\usepackage{graphicx}
\usepackage{textcomp}
\usepackage{xcolor}
\usepackage{url}
\usepackage{booktabs}
\usepackage{makecell}
\usepackage{cleveref}
\usepackage{multirow}
\def\BibTeX{{\rm B\kern-.05em{\sc i\kern-.025em b}\kern-.08em
    T\kern-.1667em\lower.7ex\hbox{E}\kern-.125emX}}
\begin{document}

\newcommand{\todol}[1]{\textsf{\textbf{\textcolor{green!55!blue}{[Lars: #1]}}}}
\newcommand{\todoj}[1]{\textsf{\textbf{\textcolor{yellow!55!red}{[Jakob: #1]}}}}
\newcommand{\todod}[1]{\textsf{\textbf{\textcolor{yellow!55!green}{[Daniel: #1]}}}}
\newcommand{\todof}[1]{\textsf{\textbf{\textcolor{cyan!55!magenta}{[Francisco: #1]}}}}

\title{This Is Taking Too Long - Investigating Time as a Proxy for Energy Consumption of LLMs}

\author{\IEEEauthorblockN{Lars Krupp}
\IEEEauthorblockA{\textit{Embedded Intelligence} \\
\textit{DFKI and RPTU}\\
Kaiserslautern, Germany \\
lars.krupp@dfki.de}
\and
\IEEEauthorblockN{Daniel Geißler}
\IEEEauthorblockA{\textit{Embedded Intelligence} \\
\textit{DFKI and RPTU}\\
Kaiserslautern, Germany \\
daniel.geissler@dfki.de}
\and
\IEEEauthorblockN{Francisco M. Calatrava-Nicolas}
\IEEEauthorblockA{\textit{Centre for Applied Autonomous Sensor Systems (AASS)} \\
\textit{Örebro University}\\
Örebro, Sweden \\
 francisco.calatrava-nicolas@oru.se}
\and
\IEEEauthorblockN{Vishal Banwari}
\IEEEauthorblockA{\textit{Embedded Intelligence} \\
\textit{DFKI and RPTU}\\
Kaiserslautern, Germany \\
vishal.banwari@dfki.de}
\and
\IEEEauthorblockN{Paul Lukowicz}
\IEEEauthorblockA{\textit{Embedded Intelligence} \\
\textit{DFKI and RPTU}\\
Kaiserslautern, Germany \\
paul.lukowicz@dfki.de}
\and
\IEEEauthorblockN{Jakob Karolus}
\IEEEauthorblockA{\textit{Embedded Intelligence} \\
\textit{DFKI and RPTU}\\
Kaiserslautern, Germany\\
jakob.karolus@dfki.de}
}

\maketitle

\begin{abstract}
The energy consumption of Large Language Models (LLMs) is raising growing concerns due to their adverse effects on environmental stability and resource use. Yet, these energy costs remain largely opaque to users, especially when models are accessed through an API — a black box in which all information depends on what providers choose to disclose. In this work, we investigate inference time measurements as a proxy to approximate the associated energy costs of API-based LLMs. We ground our approach by comparing our estimations with actual energy measurements from locally hosted equivalents. Our results show that time measurements allow us to infer GPU models for API-based LLMs, grounding our energy cost estimations. Our work aims to create means for understanding the associated energy costs of API-based LLMs, especially for end users.

\end{abstract}


\section{Introduction}

Recent progress in the field of deep learning and especially Large Language Models (LLMs) has transformed what artificial intelligence is capable of in a wide range of domains, ranging from natural language understanding to other fields such as healthcare and robotics \cite{wang2025large, haltaufderheide2024ethics}. Society experiences significant changes that move LLMs well beyond research settings into everyday use~\cite{howarth2025visited}. Large, nationally representative surveys show sharp growth in consumer adoption of generative AI tools for routine tasks, information seeking, and work/study support 
\cite{simon2025generative, mcclain202534, ofcom2024online}. In recent years, the number of proposed models has increased significantly \cite{maslej2025artificial}, further accelerating the adoption of generative AI across society. 

This technological development has been fueled by increasingly large-scale architecture, big corpora of data, and high-performance computing (HPC) infrastructure. As a result, many companies and research institutions have entered a competitive race to develop more formidable models, aiming to push the boundaries of the current state-of-the-art. 
However, this growth comes at significant computational costs, arising from both the training and inference stages of these models. The resulting energy demands, both from GPU-level power draw and system-level electricity consumption \cite{samsi2023words, strubell2020energy}, have reached new levels. Some companies, such as Google, are even constructing dedicated nuclear reactors to sustain their AI infrastructure \cite{hunt2025nuclear}. Beyond technical and economic challenges, the environmental implications of this escalating energy consumption are becoming a great concern.

Regrettably, company policies regarding the disclosure of energy consumption are often opaque, hiding true costs behind inconspicuous web user interfaces, cf. ChatGPT~\cite{openai2024gpt4technicalreport}. Closed-source models and sparse details on data center infrastructure make robust energy estimation impossible~\cite{samsi2023words, maslej2025artificial}. No direct access to weights and the model's inference configuration further prevent researchers from conducting transparent measurements of the LLM's energy footprint and, ultimately, assessing the efficiency of API-based LLM interactions.

Yet, there remains one metric that can be directly measured: task completion time. I.e., the time it takes for a model to process a single token. Consequently, computation time can prove valuable when estimating gross energy consumption of API-based LLMs. However, factors such as employed GPUs, model size and architecture, load factor, and many others still play a pivotal role and can drastically change computation times. Grounding results via locally deployed models with predefined options is crucial to consolidating time-based estimations.

In this work, we investigate the potential of inference completion time as a robust proxy for estimating energy consumption in large language models (LLMs). To this end, we present a benchmark designed to analyze the relationship between inference time and energy usage under controlled experimental conditions for both locally hosted and API-based models. We focus on Mistral models~\cite{mistral2025mistral}, as they are available in both open-weight and API-accessible versions. The goal of this work is to evaluate whether inference time can serve as an effective proxy for energy estimation in API-based Mistral models and consolidate our time-based estimates via locally measured computation times under representative conditions. Our code and evaluation results are available on GitHub\footnote{\url{https://github.com/DFKIEI/LLM-Energy}}.

While our work is not the first to propose API-based energy estimation, cf.~\cite{jegham2025hungry}, our approach grounds API-based energy estimation through local measurements using the same model run on representative computing infrastructure. Our results show that inference time measurements on API-based LLMs, grounded by equivalent measurements on local models, allow us to infer the GPU configurations used on external servers. Our work provides evidence of a direct correspondence between API-based large language models and their local counterparts in terms of computation time and, ultimately, energy consumption.


\section{Related Work}
Starting with the inception of the attention mechanism~\cite{vaswani2017attention}, large language models, such as ChatGPT, have been gaining traction, leading to fierce competition between different model families like OpenAI~\cite{openai2024gpt4technicalreport}, Llama~\cite{dubey2024llama}, deepseek~\cite{liu2024deepseek}, qwen~\cite{yang2025qwen3}, and mistral~\cite{mistral2024nemo}. However, outside of competing on different benchmarks, there is ongoing discussion about the topic of open-source and proprietary models. While proprietary models often show superior performance on benchmarks~\cite{white2024livebench}, their inner workings frequently remain opaque, in contrast to open-source LLMs. Especially regarding the energy consumption of proprietary models, information is sparse and often unreliable, if available at all. 

The energy footprint of LLMs spans both training and inference phases and constitutes one of the most pressing sustainability challenges in AI. 
Recent large-scale systems such as LLaMA-65B are estimated to consume around $5\times10^5$ kWh during training~\cite{strubell2020energy,patterson2021carbon}. 
Despite significant gains from modern datacenter efficiency and AI-optimized hardware, such runs can still emit tens to hundreds of tons of $CO_2$ when powered by fossil-based energy~\cite{henderson2020towards,patterson2021carbon}. 
Advances in model sparsity (e.g., mixture-of-experts), efficient accelerator architectures, and low-carbon datacenters can reduce energy intensity by one to three orders of magnitude~\cite{patterson2021carbon,zhao2022green}. 

For inference, even though a single LLM query may consume only 0.3 to 1\,Wh, the cumulative demand of millions of daily requests may scale to annual petawatt-hour levels by 2026~\cite{ozcan2025quantifying,desislavov2023trends}.
Inefficiencies from idle GPU power draw, throughput limits, and non-proportional energy scaling further amplify emissions, whereas techniques such as quantization, batching, and dynamic scheduling can mitigate these effects~\cite{ozcan2025quantifying,wu2022sustainable}.

Accurately assessing the energy footprint of large language models remains challenging. Recent tools such as CarbonTracker~\cite{anthony2020carbontracker} and CodeCarbon~\cite{benoit_courty_2024_11171501} aim to provide real-time monitoring of power consumption and $CO_2$ emissions by integrating hardware-level energy data with regional grid carbon intensity. 

Recent studies have begun to quantify the energy and environmental costs associated with both the training and evaluation stages of deep learning systems. On the training side, Geißler et al.~\cite{geissler2024power} analyzed how hyperparameters can influence the energy demand of neural networks, showing that suboptimal hyperparameter choice can increase the energy consumption even when achieving similar accuracy. As model scales have grown, these concerns have extended from conventional deep networks to large language models (LLMs). Fernandez et al. \cite{fernandez2025energy} presented a comprehensive study of the energy implications of LLM inference, examining how model architecture, decoding strategies, and software frameworks affect power use. Jegham et al. \cite{jegham2025hungry} further introduced time-based estimation methods to assess the energy consumption of API-served models, yet lacking grounding through actual measurements. Here, Krupp et al. \cite{krupp2025promoting} depict an approach for benchmarking the energy consumption of locally run web agents and expose the complexity of estimating the energy cost for web agents that use API-based LLMs.

\section{Methodology}
\label{sec:methodology}
Building upon these works, we propose a methodology that leverages locally executed models to inform the energy estimation of API-based systems, using inference computation time as a common factor for energy consumption across both setups. Our core \textbf{assumption} is that when computational hardware operates at or near optimal utilization, its power draw remains approximately constant, allowing total energy consumption to be estimated as a function of execution time. However, since the exact hardware specifications for API-based models are undisclosed, we additionally benchmark models locally across different GPU architectures to establish a plausible mapping of energy estimates to GPU fleets. We then compare these results with the inference completion times measured from API-based models. 

Our benchmark consists of synthetic prompts (generated with llama3-70b) from four input–output configurations (covering technical, creative, educational, and business-oriented tasks with injected topics) that differ in sequence length (short = 2,048 tokens, long = 8,192 tokens). An example educational prompt may look like this: "Provide a thorough explanation of advanced mathematics".

We selected the LLMs based on their availability in both open-weight and API-based variants, as well as their model size. These criteria enabled us to assess the accuracy of our time-based proxy energy estimation approach more reliably by testing both models locally and through their API, while ensuring that the models remain lightweight enough to fit on a single GPU when tested locally. Specifically, we used \textit{Mistral-7B-Instruct-v0.3} (Local), equivalent to \textit{Open-Mistral-7B} (API), and \textit{Mistral-NeMo-Instruct-2407} (Local), equivalent to \textit{Open-Mistral-NeMo} (API).

\subsection{Experiment Protocol}
\label{sub_sec:experiment_protocol}
To ensure reproducibility and consistency, all tests are conducted following a structured experimental protocol. This applies to both locally executed runs as well as runs initiated on API-based models.
Each run processes a fixed sequence of 100 synthetic prompts from our benchmark. 
This deterministic ordering eliminates sampling variance, ensuring that all models and hardware platforms receive identical workloads.

Execution proceeds in structured passes, where each configuration is executed repeatedly to ensure statistical reliability. 
For every configuration, the full benchmark is run independently ten times, allowing mean and standard deviation values to be computed for all runtime and energy metrics. In particular, for API-based models, we schedule runs throughout the day to capture variances due to times of high load.
For each locally executed run, all prompts are processed in parallel batches of eight to ensure deployment with sufficient GPU utilization. Idle cycles between queries are minimized to ensure consistent energy-per-token comparability. 

\subsection{Local Hardware Architecture Selection}
\label{sec:hardware_architecture}
To assess the architectural impact on runtime and energy efficiency, local experiments are performed across six NVIDIA GPUs representing two generations (Ampere, Hopper) and form factors (SXM, PCIe): A100-40GB, A100-80GB, A100-PCI, H100, H100-PCI, and H200. 
These devices differ markedly in thermal design power (TDP) and optimal sustained load range as shown in \Cref{tab:gpu_power_optimal}.

The comparison focuses on the divergence between server-grade SXM modules and PCIe variants (A100-PCI, H100-PCI). 
Server versions benefit from higher TDP ceilings, external cooling, and high-bandwidth NVLink interconnects, whereas PCIe versions have less power and much larger auxiliary losses, such as power conversions and cooling, which are addressed on-chip.

Public sources do not disclose the exact GPU fleet serving Mistral APIs. However, Mistral reports training on H100 GPUs~\cite{nvidia_mistral_h100}, and MLPerf v4.1 shows H100/H200-class accelerators dominate datacenter LLM inference~\cite{nvidia_mlperf_v41}. Since Mistral models are offered via major clouds~\cite{aws_bedrock_mistral}, H100/H200-class GPUs are most likely used for API-hosted deployments. 

\subsection{Local LLM Initialization}
The evaluated models, Mistral-7B-v0.3 and Mistral-NeMo-Instruct-2407 (12B), are executed in FP16 precision with TF32 acceleration on Ampere and Hopper GPUs, using device-map="auto" for balanced memory allocation. KV-caching and left-padding are applied to optimize batch consistency and reduce redundant computation.

To ensure comparability, temperature is fixed at 0.7 across all runs, for both local and API-based models, mimicking real use cases. 
Random seeds (42) are synchronized across Python, NumPy, PyTorch, and CUDA. Prompts are passed directly to the model without system instructions, ensuring that runtime and energy variations result solely from model architecture, precision, and decoding configuration.

\subsection{Local Energy Tracking}
\label{sub_sec:local_energy_tracking}
Energy consumption is recorded with CarbonTracker \cite{anthony2020carbontracker}, which samples GPU power via NVIDIA System Management Interface (SMI) and integrates the full board-level energy.
In contrast, TDP reflects only the chip’s power envelope, serving as a theoretical lower bound. 
The difference between the two indicates the additional power drawn by auxiliary components.

This gap is most evident for PCIe GPUs, where less efficient power delivery and higher conversion losses cause CarbonTracker values to exceed TDP estimates, while server-grade GPUs show closer alignment. 
Comparing both metrics thus reveals how chip-level specifications underestimate real-world energy use, especially in PCIe configurations.

\subsection{API-Based Energy Estimation}
\label{subsec:metric_cal}
As outlined in our experiment protocol, we performed $N=10$ independent executions on both a \textit{local open-weights deployment} and an \textit{API-based deployment}. For each local run $i \in \{1, 2, \dots, N\}$, we tracked energy consumption during code execution. The resulting energy consumption is denoted as $E_i^{\text{loc}}$, and the wall-clock duration of each run as $T_i^{\text{loc}}$. Consequently, \textit{average power} draw for a local run can be calculated as follows: $P_i^{\text{loc}} = \frac{E_i^{\text{loc}}}{T_i^{\text{loc}}}$, averaging over all runs yields ${P}^{\text{loc}}$.



Based on this local power calculation, we can predict the \textit{energy consumption} of the equivalent API-based model, if run on a comparable computing setup:
\begin{equation}
  \widehat{E}^{\text{api}}
  \;=\;
  {P}^{\text{loc}} \cdot \bar{T}^{\text{api}},
  \label{eq:eapi}
\end{equation}
where $\bar{T}^{\text{api}}$ denotes the mean API execution time over all API runs.

To allow comparability between runs and models\footnote{Output token count was not consistent across models.}, we normalized our results by token count. Consequently, for each run, we normalized both energy $E_i$ and time $T_i$ by the token count of this run $N^{(i)}_{tokens}$. This yields mean energy per token $\bar{E}_{token}$ and mean inference time per token $\bar{T}_{token}$.

\section{Results}
This section presents the evaluation of energy consumption for large language models (LLMs) from the Mistral family, executed both locally and through API-based services.
We begin by analyzing the computational time required to complete the benchmark under both scenarios (local vs. API), as it forms the basis for subsequent energy estimation. Next, we present the energy consumption analysis, reporting both the total energy and energy per token (see Section \ref{subsec:metric_cal}). Finally, we validate the energy predictions for the local setup by comparing them with estimates derived from the Thermal Design Power (TDP) parameter of each GPU.

\subsection{Benchmarking Computation Time}
\label{sub_sec:benchmarking_time}



   \begin{table}[t]
    \centering
    \caption{TDP and Optimal Work Load Range}
    \label{tab:gpu_power_optimal}
    \begin{tabular}{lccc}
    \toprule
    GPU & VRAM&\makecell{Thermal Design\\Power (TDP) [W]} & \makecell{Optimal TDP\\Load Range [W]}\\
    \midrule
    A100-40GB  & 40 GB  & 400 & 340–380 \\
    A100-80GB  & 80 GB  & 400 & 340–380 \\
    A100-PCI   & 40 GB  & 250 & 213–238 \\
    H100       & 80 GB  & 700 & 595–665 \\
    H100-PCI   & 94 GB  & 400 & 340–380 \\
    H200       & 140 GB & 700 & 595–665 \\
    \bottomrule
    \end{tabular}
\end{table}

\begin{table}[h!]
    \centering
    \caption{Table depicting mean and standard deviation of time (s) to compute the benchmark. GPUs are ordered chronologically by release date.}
    \label{tab:runtimes}
    \begin{tabular}{cccc}
    \toprule
   Model  &    Type     &    GPU    &  mean $\pm$ sd \\
          \midrule
\multirow{8}{*}{Mistral-7B}    & \multirow{6}{*}{Local}   & A100-40GB & 2447. $\pm$ 8.85\\
    &    & A100-80GB & 2308. $\pm$ 8.36\\
    &    & A100-PCI  & 2525. $\pm$ 14.7\\ 
    &    & H100      & 1662. $\pm$ 22.5\\ 
    &    & H100-PCI  & 1820. $\pm$ 9.24\\
    &    & H200      & 1559. $\pm$ 4.72\\
    \cmidrule{2-4}
    & Free-API       & - & 718.  $\pm$ 43.8\\ 
    & Paid-API       & - & 727.  $\pm$ 28.8\\
\midrule

\multirow{8}{*}{Mistral-NeMo}    & \multirow{6}{*}{Local} & A100-40GB & 3865. $\pm$ 8.13\\
    &  & A100-80GB & 3612. $\pm$ 14.0\\ 
    &  & A100-PCI  & 4003. $\pm$ 16.4\\ 
    &  & H100      & 2140. $\pm$ 23.1\\ 
    &  & H100-PCI  & 2353. $\pm$ 11.9\\ 
    &  & H200      & 2018. $\pm$ 8.17\\
    \cmidrule{2-4}
    & Free-API & - & 1260. $\pm$ 161.\\  
    & Paid-API & - & 1159. $\pm$ 31.2\\ 
\bottomrule
    \end{tabular}
\end{table}

\Cref{tab:runtimes} shows the runtime on our proposed benchmark over ten runs for both scenarios (local and API). The comparison between local and API-based executions highlights a substantial difference in computational efficiency. Both free and paid API run the benchmark faster than any local GPU configuration. Compared to the best-performing local setup, the API execution achieves a reduction in computation time of approximately 54\% for Mistral-7B and approximately 43\% for Mistral-NeMo, indicating differences in the setup, such as system prompts or parallelization. The marginal difference between the free and paid APIs (on the order of a few seconds, within one standard deviation) suggests that both rely on similar backend resources, with variations likely due to server load or queueing effects. In contrast, local executions display longer but more predictable runtimes (smaller standard deviation), where performance depends on the underlying hardware. Within the local setup, a clear hierarchy emerges: newer GPUs (H100 and H200) consistently complete the benchmark faster than the older Ampere series models.
\begin{figure}[t!]
    \centering
    \includegraphics[width=\linewidth]{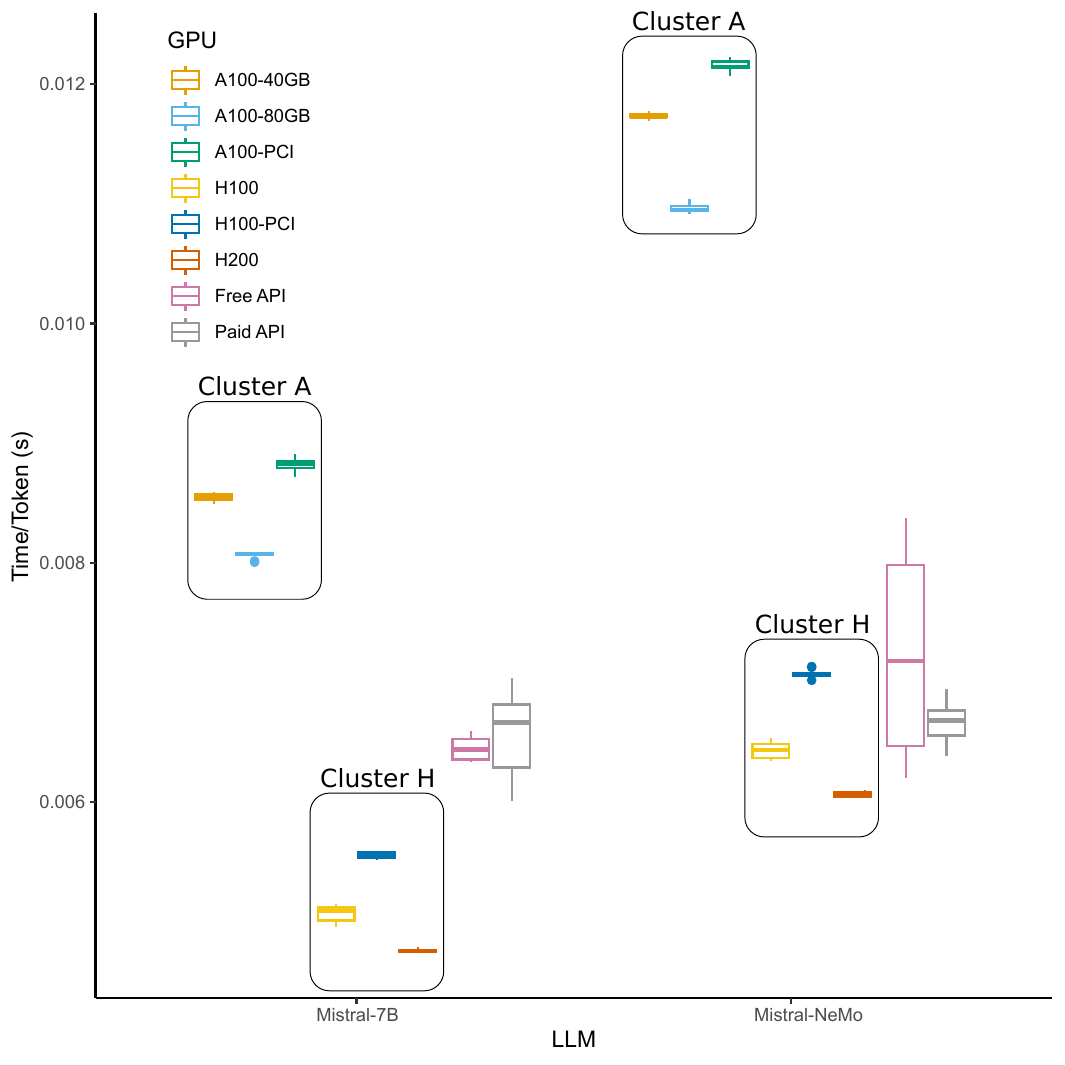}
    \caption{Boxplot showing the computation time per token $\bar{T}_{token}$ in seconds for running the same benchmark on both models across different local GPUs and API executions. Cluster A includes A100 GPUs, while Cluster H groups H100 and H200 GPUs.}
    \label{fig:energy_per_time}
\end{figure}

In addition, we offer a fine-grained comparison using the average time per token. 
As seen in \Cref{fig:energy_per_time}, the $\bar{T}_{token}$ of the different configurations are clustered by GPU generation for local runs. For both LLMs, we observe two local clusters (see \Cref{fig:energy_per_time}), the A-Cluster and the H-Cluster. The API-based $\bar{T}_{token}$ for Mistral-NeMo are close to the H-cluster, however, this is not the case for Mistral-7B, where runtimes are slightly higher than for the H-Cluster GPUs. \Cref{sub_sec:estimating_energy} puts these results into perspective regarding energy consumption.


\subsection{Translating Time to Energy}

In this section, we present the results related to the use of time in energy estimation for the API models (c.f. \Cref{subsec:metric_cal}). In addition, we compare this value with the energy consumption based on the Thermal Design Power (TDP) for each GPU. We use an average sustained power equal to 0.9 × TDP as a conservative estimate where cooling cost and energy losses through heat radiation are not included. Empirically grounded estimates of actual energy use during high-throughput inference are reported between 85\,\% and 95\,\%, cf.~\cite{samsi2023words,nvidia_mistral_h100,nvidia2020a100}.
Based on these findings, \Cref{tab:gpu_power_optimal} reports the optimal TDP load values adopted in our analysis.

Applying this information to the calculated runtimes, we can calculate a lower bound for the energy consumption. As shown in \Cref{tab:gpu_power_optimal_grouped}, the differences in energy consumption between TDP and local measurements can vary widely, depending on the GPU architecture (SXM or PCIe), as outlined in \Cref{sec:hardware_architecture}.

\begin{table*}[h!]
    \centering
    \caption{Average sustained power and energy use estimated via TDP vs. locally measured (Carbon Tracker). Each super-level group per token and total energy.}
    \label{tab:gpu_power_optimal_grouped}
    \begin{tabular}{l l c cc cc c}
        \toprule
        \multirow{2}{*}{Model} & \multirow{2}{*}{GPU} & \multirow{2}{*}{Total tokens} 
        & \multicolumn{2}{c}{\textbf{TDP (calculated)}} 
        & \multicolumn{2}{c}{\textbf{Carbon Tracker (measured)}} 
        & \multirow{2}{*}{\makecell{Difference\\(\% of local)\\mean [\%]}} \\
        \cmidrule(lr){4-5}\cmidrule(lr){6-7}
        & & 
        & \makecell{Per token (mWh)} 
        & \makecell{Total (Wh)} 
        & \makecell{Per token (mWh)} 
        & \makecell{Total (Wh)} 
        & \\
        \midrule
        \multirow{6}{*}{Mistral-7B}   & A100-40GB & 286219 & 0.86 $\pm$ 0.003 & 245. $\pm$ 0.885 & 1.06 $\pm$ 0.023 & 303. $\pm$ 6.54 & 19 \\
           & A100-80GB & 286219 & 0.81 $\pm$ 0.003 & 231. $\pm$ 0.836 & 1.10 $\pm$ 0.024 & 314. $\pm$ 6.75 & 26 \\
           & A100-PCI  & 286219 & 0.55 $\pm$ 0.003 & 158. $\pm$ 0.921 & 0.949 $\pm$ 0.007 & 272. $\pm$ 1.95 & 42 \\
           & H100      & 327674 & 0.88 $\pm$ 0.012 & 291. $\pm$ 3.94  & 1.07 $\pm$ 0.016 & 350. $\pm$ 5.37 & 16 \\
           & H100-PCI  & 327674 & 0.56 $\pm$ 0.003 & 182. $\pm$ 0.924 & 0.904 $\pm$ 0.009 & 296. $\pm$ 2.89 & 39 \\
           & H200      & 327674 & 0.83 $\pm$ 0.003 & 273. $\pm$ 0.827 & 1.04 $\pm$ 0.011 & 340. $\pm$ 3.54 & 20 \\
        \midrule
        \multirow{6}{*}{Mistral-NeMo} & A100-40GB & 329345 & 1.15 $\pm$ 0.002 & 387. $\pm$ 0.813 & 1.47 $\pm$ 0.027 & 485. $\pm$ 8.75 & 20 \\
         & A100-80GB & 329345 & 1.10 $\pm$ 0.004 & 361. $\pm$ 1.40  & 1.54 $\pm$ 0.015 & 507. $\pm$ 4.83 & 29 \\
         & A100-PCI  & 329345 & 0.75 $\pm$ 0.003 & 250. $\pm$ 1.02  & 1.32 $\pm$ 0.007 & 435. $\pm$ 2.23 & 43 \\
         & H100      & 332860 & 1.13 $\pm$ 0.012 & 375. $\pm$ 4.04  & 1.37 $\pm$ 0.019 & 456. $\pm$ 6.28 & 18 \\
         & H100-PCI  & 332860 & 0.71 $\pm$ 0.004 & 235. $\pm$ 1.19  & 1.17 $\pm$ 0.013 & 388. $\pm$ 4.25 & 39 \\
         & H200      & 332860 & 1.06 $\pm$ 0.004 & 353. $\pm$ 1.43  & 1.33 $\pm$ 0.021 & 444. $\pm$ 7.00 & 20 \\
        \bottomrule
    \end{tabular}
\end{table*}

\subsection{Estimating the Energy Consumption for API-Based LLMs}
\label{sub_sec:estimating_energy}
By contrasting the measured time per Token $\bar{T}_{token}$ of our GPU setups with the results from the API-based models, we can pinpoint likely GPU setups for the given LLMs. We do note that, while $\bar{T}_{token}$ might be comparable, we cannot make any evidence-based statements on the actual GPU setups behind the API-based models. As such, this comparison is based on reports about compute clusters and our confirming findings through inference time measurements regarding those GPU setups. Likewise, \Cref{fig:energy_per_time} reveals that the H100-PCI is a good representative setup for both Mistral models.



Finally, we calculated the estimate of mean energy consumption for the API-based models $\widehat{E}^{\text{api}}$ as seen in \Cref{eq:eapi} using the average number of tokens computed by the paid and free API and energy per token results for both TDP calculation and local measurements. For Mistral-7B and for Mistral-NeMo, the results are shown in \Cref{tab:mistral_estimation} showcasing differing energy consumptions. Note that TDP calculations are likely underestimated, cf.~\Cref{sub_sec:local_energy_tracking}. We will discuss additional factors for these results in the next section. 


\begin{table}[t]
\scriptsize
    \centering
        \caption{Comparison between measured and calculated total energy costs for the API version (free and paid) of Mistral-7B and Mistral-NeMo.}
    \label{tab:mistral_estimation}
    \begin{tabular}{lllrr}
    \toprule
   Model & API cost& &  \makecell{Number of Tokens\\ }& \makecell{H100-PCI\\Total energy (Wh)\\} \\
\midrule
Mistral-7B & free & M     & 111272. $\pm$ 6351. &  100.60 $\pm$ 6.00\\
Mistral-7B & free  & C    & 111272. $\pm$ 6351.  & 62.31 $\pm$ 3.56 \\
[1ex]
Mistral-7B & paid & M & 110882. $\pm$ 5922. & 100.23 $\pm$ 5.60\\
Mistral-7B & paid & C & 110882. $\pm$ 5922.  & 62.09 $\pm$ 3.32 \\
\midrule
Mistral-NeMo & free & M &173743  $\pm$ 4967. &  203.28 $\pm$ 5.81 \\
Mistral-NeMo & free & C & 173743  $\pm$ 4967. &  123.36 $\pm$ 3.53 \\
[1ex]
Mistral-NeMo & paid & M & 174496. $\pm$ 5815. & 204.16 $\pm$ 6.80 \\
Mistral-NeMo & paid & C & 174496. $\pm$ 5815. &  123.89 $\pm$ 4.13 \\
\bottomrule
    \end{tabular}
    \vspace{1mm}
    \footnotesize \\M = measured; C = calculated.
\end{table}

\section{Discussion}

Our results confirmed the necessity for a strict experiment protocol. Deviations in deployment-specific factors, such as variations in system prompts and minor model version differences between API and open-source releases, may easily lead to skewed time measurements as depicted in the comparison between \Cref{tab:gpu_power_optimal_grouped} and \Cref{tab:mistral_estimation}. Here, the API and local executions produce different numbers of tokens, making raw total energy values misleading across settings. Consequently, normalizing energy values given the amount of processed tokens is essential for accurate results.



\Cref{fig:energy_per_time} shows these normalized values for API-based and local models. In particular, it highlights a close alignment between the API-based deployment and Cluster H (local deployment equipped with NVIDIA Hopper series GPUs) for Mistral-NeMo. However, we note that this does not imply that both deployments necessarily consume the same energy per token. Server-side optimization techniques, such as tensor or pipeline parallelism, may be applied to reduce latency at the expense of higher power consumption. For instance, hosting Mistral-7B on multiple GPUs shortens the runtime but proportionally increases total energy consumption per second.


Nonetheless, we know that most data centers are using a mixture of Nvidia Ampere and Nvidia Hopper series GPUs. Further, our local time per token results are within one standard deviation of the API-based results. As we select models that fit into a single GPU, it is reasonable to conclude that the two measurements directly correspond, ultimately confirming our hypothesis that inference time per token is a reasonable proxy for energy estimation in API-based LLMs.




\subsection{Limitations}
This work relies on the assumption that the time we measure for the API-based models is dominated by GPU computation time, with other contributions being negligible. 
This assumption allowed us to reduce the complexity of our evaluation, such as only one GPU and a single batch size. 
In practice, however, real-world deployments may introduce additional sources of latency and energy consumption that are not directly observable in our proposed experimental setup. These include multi-GPU execution, as well as tensor, pipeline, and expert parallelism strategies, which can change the runtime–energy trade-off (e.g., reducing latency at the cost of increased power draw).
While our approach therefore provides a reasonable gross estimate of the energy consumption of LLMs accessed through an API, a more fine-grained characterization of these effects can improve accuracy. As future work, systematic ablation studies, including parallelism strategies, hardware configurations, and batch sizes, are necessary to quantify the impact of these factors on both runtime and energy use.

\section{Conclusion}
In this work, we investigated the potential of inference time as a proxy for estimating the energy consumption of API-based language models. Our results show that time per token can be used to imply the type of GPU configurations used behind API-based models. Likewise, gross energy estimations for these particular models are possible, providing researchers with a means to judge the energy efficiency of API-based models.

\section{Acknowledgments}
This work is funded and supported by the EU’s Horizon Europe research and innovation program through the ”SustainML” project (101070408), the Carl Zeiss Foundation through the project ”Sustainable Embedded AI”, the German BMFTR through the "Cross-Act" project (01IW25001) and the Wallenberg AI, Autonomous Systems and Software Program (WASP) under the Knut and Alice Wallenberg Foundation.

\bibliographystyle{IEEEtran}   
\bibliography{OurReferences}
\end{document}